# Analysis of Mumbai Floods in recent Years with Crowdsourced Data

Shrabani Sailaja Tripathy[1], Sautrik Chaudhuri[2], Raghu Murtugudde[1], Vedant Mharte[3], Dulari Parmar[4], Manasi Pinto[4], P.E. Zope[2], Vishal Dixit[1], Subimal Ghosh[1,2,*]

[1]Interdisciplinary Program in Climate Studies, Indian Institute of Technology Bombay, Powai, Mumbai-400076, India.
[2]Department of Civil Engineering, Indian Institute of Technology Bombay, Powai, Mumbai-400076, India.
[3]Electronics and Telecommunication Engineering, Vidyalankar Institute of Technology, Mumbai 400 037, India
[4]Youth for Unity and Voluntary Action (YUVA), Kharghar, Navi Mumbai, 410201, India

*Corresponding author, subimal@iitb.ac.in

# Abstract

Mumbai, a densely populated city, experiences frequent extreme rainfall events leading to floods and waterlogging. However, the lack of real-time flood monitoring and detailed past flooding data limits the scientific analysis to extreme rainfall assessment. To address this, we explore the usability of crowdsourced data for identifying flood hotspots and extracting reliable flood information from the past. Through an automated program, we filter and retrieve flood-related data from Twitter, using location information to generate flood maps for past heavy rainfall events. The validity of the retrieved data is confirmed by comparing it with volunteered geographic information (VGI) which is more accurate but less abundant. In the absence of direct flood information, Twitter data is cross-verified with the Height above the Nearest Drainage (HAND) map, which serves as a proxy for elevation. Interestingly, while extreme rainfall events are increasing in frequency, recent Twitter-based information shows a decrease in flood reporting, attributed to effective mitigation measures implemented at various flood hotspots. Local surveys support this finding and highlight measures such as underground storage tanks and pumping stations that have reduced flood severity. Our study demonstrates the value of crowdsourced data in identifying urban flood hotspots and its potential for real-time flood monitoring and forecasting. This approach can be adapted for data-sparse urban regions to generate location-specific warnings, contributing to improved early warnings and mitigating the impact on lives and property.

# 1. Introduction

Every year flooding causes major social and economic damage and affects millions of people around the globe. The magnitude and frequency of floods are increasing in most parts of the globe at an alarming rate as a result of the complex interaction of hydrology, human involvement and climate. Depending on the mechanism of generation, spatial and temporal scales, floods are categorised into various classes (Nied et al., 2014). Urban flooding is a type of pluvial flood that occurs due to saturating drainage capacity in an urban area. Various drivers such as torrential rain, river overflow, coastal storm surge, sea-level rise, flash floods, groundwater seepage, sewer overflow, lack of permeability, and lack of city management can cause these floods. Global warming has increased extreme precipitation, which gets further intensified in urban regions, resulting in severe flash/ pluvial floods due to low permeability (Dodman et al., 2022).

Urban flooding results in various direct and indirect impacts, socioeconomic ramifications, and risks to human life and health, highlighting the interdependencies of infrastructure systems in urban spaces (Kim et al., 2017; Rosenzweig et al., 2021). An effective high-resolution flood forecasting/ early warning system reduces the risk due to urban flooding. The performance of a flood forecasting system depends on atmospheric and hydrologic factors and human management. The hydrological factors include real-time flood monitoring that provides initial conditions for the next few hours to days, flood modeling, and reliable long-term validation data of water logging at different flood hotspots within a city. In spite of the identification of these factors, monitoring urban floods at high spatial and temporal resolution remains challenging. Monitoring stations are often sparsely located, and hence may not be a good representative of the variability in the large urban region. Further, there is a lack of a participatory framework involving stakeholders, i.e., the citizens, to gather specific local knowledge regarding floods (Gebremedhin et al., 2020) in developing high-resolution models and validating forecasts.

The dense population in urban areas can be leveraged as an advantage by utilizing crowdsourced data for the purpose of flood monitoring and forecasting. With high population density, there is ample availability of communication systems such as cell phones, social media, traffic cameras, webcams, etc., which can be used to gather real-time and location-specific data (Helmrich et al., 2021; Tavra et al., 2021). Crowdsourcing is an opportunistic collection of data from otherwise autonomous dispersed sources. This helps gather large amounts of real-time data cost effectively. Various crowdsourced data sources include social media like Twitter, Facebook, and Instagram (Arthur et al., 2018; Bruijn et al., 2019; Bruijn et al., 2020; Helmrich et al., 2021; Smith et al., 2017; Songchon et al., 2021; R. Wang et al., 2018), Mobile apps, Volunteered Geographical Information (VGI) (Degrossi et al., 2014; Goodchild, 2007), public webcams, various personal sensors, etc. Recent studies have enabled communities in gathering and providing data to help in the disaster management and mitigation of various natural disasters; earthquakes (Bondár et al., 2020; Minson et al., 2015), floods (Gebremedhin et al., 2020; Havas et al., 2017; Helmrich et al., 2021), forest fires (Bielski et al., 2017; Çolak & Sunar, 2020; Oliveira et al., 2019), heat waves (Roberts, 2022; Venter et al., 2021), and such. These crowdsourced data have been used to identify flood events, collect flood and rainfall data, and promote preparedness and adaptation in disaster risk management (Chen et al., 2022; Hultquist & Cervone, 2020; Paul et al., 2018; Y. Wang et al., 2018). The common datasets used by previous studies include public webcams, social media, various sensors, mobile applications, etc. (Blumberg et al., 2015; Degrossi et al., 2014; Gebremedhin & Basco-carrera, 2020; Sy, et al., 2019; Witherow et al., 2018).

Real-time availability and dense networking are the major advantages of crowdsourced datasets. For urban flood forecasting, crowdsourced data can be used at different stages of a flood (See, 2019):

a.  Before the flood: Flood forecasting system and model development.
b.  During the flood: Flood identification and warning generation for an ongoing flood event
c.  After the flood: Damage and impact assessment, model improvements and skill assessment.

Out of the various social media platforms like Facebook, Twitter, Instagram, we have considered Twitter as the data source for our study because of the large data availability and easy accessibility. The Application Programming Interface (API) of websites allows the user to access and retrieve information using any programming platform. Nevertheless, the quality of information retrieved remains a major concern that needs to be taken care of by quality controlling techniques (Flanagin & Metzger, 2008; Helmrich et al., 2021; Kosmala et al., 2016; Songchon et al., 2021). Though some studies have used this concept for different cities, this concept is yet to be explored for a highly populated megacity like Mumbai, India.

In this study, we demonstrate the feasibility of using Twitter and VGI information for flood data collection and monitoring for Mumbai City. Useful information was extracted from the crowdsourced data by filtering out duplicate and irrelevant tweets. The data was obtained through the Twitter API and validated against past events by cross-referencing them with observed extreme rain events. Additionally, to validate the conclusions drawn from the crowdsourced information, a survey was conducted.

## 1.1. Study Area

Mumbai, a megacity in India, extending between 18.00-19.20ºN and 72.00-73.00ºE is the case study area. Greater Mumbai covers an area of 603 km$^2$ (233 sq. mi), consisting of the Mumbai City and Mumbai Suburban districts, which extend from Colaba in the south to Mulund and Dahisar in the north, and Mankhurd in the east. It is further divided into 24 municipal wards and 227 co-operators wards (Figure 1). The city is popularly known as the financial capital of India and has a current population of approximately 20 million.

The city receives an annual average rainfall of more than 2000 mm (Hallegatte et al., 2010). Rainfall mostly occurs during the four months of monsoon, June to September, with a number of high intensity extreme events. The primary causes behind the torrential rainfall are orography and uncontrolled urbanization (Shastri et al., 2017). The geographic location of any place plays a major role in the climate of that location. Mumbai is situated on the windward side of India's Western Ghats mountain ranges. The south-westerly monsoon flows from the Arabian Sea bringing enormous moisture that produces heavy rain due to orography. The varying elevation level and urbanization, add to the intensity of heavy rainfall, generating frequent floods in the city. Almost every year, the city suffers from multiple water logging and flood situations at various locations (Shastri et al., 2017). Apart from huge losses, this hampers the everyday lives of the citizen.

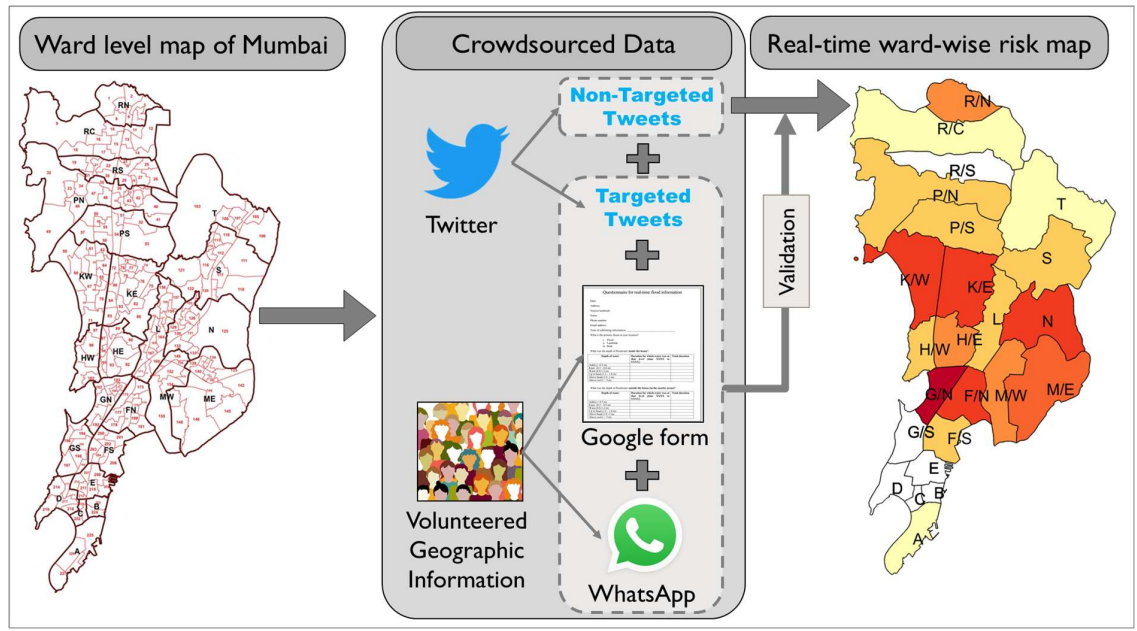

*Figure 1. 24 municipal wards and 227 co-operators wards of Mumbai; Data collection and methodology followed to generate real-time ward-wise risk map*

## 2. Method

### 2.1. Data retrieval: Twitter

Social media platforms like Facebook, Instagram, Twitter, etc. have been extensively used for flood identification and model development as indicated by literature (Arthur et al., 2018; Bruijn et al., 2019; de Bruijn et al., 2020; Helmrich et al., 2021; Smith et al., 2017; Songchon et al., 2021; R. Wang et al., 2018). For our study, we used the Twitter platform and volunteered geographic information (VGI) to gather data (Figure 1). Twitter is preferred over other social media platforms as most of the Twitter accounts are public, which allows easy access to the information. The size of tweets is limited to 280 characters, often making them crisper with sufficient details. These tweets are retrieved through Twitter Streaming API, which provides access to a high volume of tweets to retrieve and analyse the information. From an academic account 10,000,000 numbers of tweets can be retrieved per month. We have developed a Python based (tweepy library)(Joshua Roesslein, 2009) automated program to retrieve data from Twitter and store it in a .csv format. At one-time step, a maximum of 100 tweets can be retrieved (1 request per second). This code contains various filters so that a focused search can be performed for the desired information. The details of the filters used are discussed in the following subsection. The developed code is generic and can be used for any preferred data retrieval or location.

### 2.1.1. Data collection

#### Targeted and Non-Targeted Tweets

The data gathered on Twitter is classified into two groups based on the intention of the tweet. Targeted tweets are the volunteered tweets that are done following the requested format by our official Twitter account (https://twitter.com/ClimateIITB?s=20) . These tweets can be used as the absolute ground truth, as these tweets are made on purpose to provide us the flood depth information. As a part of a campaign started under this research project, we have requested users to tweet regarding the flood information in their area using a particular format and hashtag. The specified format allows easier and more efficient data retrieval through predefined filters. However, these tweets tend to be very low in number and cannot be used as a sole source of finer-scale data. The non-targeted tweets are the normal tweets put out by people without the intention to provide us with the information. These tweets need to be analysed further to gather useful information. The process of data retrieval from non-targeted tweets is discussed in the following sections.

### 2.1.2. Filters

Effective filtering of tweets is necessary to gather useful information and ensure quality (de Bruijn et al., 2020; Degrossi et al., 2014; Flanagin & Metzger, 2008). During the data retrieval, we used various filters to ensure useful information. The filters considered in this study are:

- Language filter: English
- No retweets
- Duration (Start and end time)
- Time step

The Python based code uses this information as input to retrieve tweets and store them in a tabular (.csv) form. To gather tweets that concern our needs, we have shortlisted the most popular keywords used by account holders to tweet regarding floods and water logging in Mumbai. The three most popular hashtags used are:

i.   #mumbairains
ii.  #mumbairain
iii. #mumbaiflood

For July 2022, we have retrieved all the tweets using all these hashtags along with all the filtered tweets based on all the keywords. It is seen that #mumbairains is the most used keyword (Supplementary figure 1) and all other keywords are mostly used in the same tweets. In order to gather maximum information and to reduce the chances of duplications we have only considered #mumbairains as the keyword.

## 2.2. Location identification and Assigning flood risk proxy

Geotagged tweets have the exact location information and hence, they are the most useful tweets. However, only 2-3% of the total tweets are geotagged; most of the retrieved tweets may not be geotagged and hence directly getting the location information of each tweet is not possible. Hence, the text of the tweets was analysed to recover the location information of a flood or waterlogging (Supplementary table 1). In order to accomplish this, we created a directory for 442 possible location names from Mumbai with their ward and corresponding latitude - longitude. The program matches the location name from the tweets and assigns a value of one to the corresponding wards. A single tweet may be not reliable, but a large number of unrelated tweets mentioning a flood in the same region can provide genuine information (Bruijn et al., 2019). Hence, the higher the number of tweets showing flood in an area (ward), the worse is the flood situation.

## 2.3. Volunteered Geographic Information (VGI)

Ensuring the quality of crowdsourced data poses a significant challenge, particularly when information is sourced from platforms like Twitter. In order to maintain the accuracy and reliability of the data gathered, it was imperative for us to verify the information received. However, there is no authentic source of flood depth information available for Mumbai, to the best of our knowledge. Hence, we have used the information provided by volunteers at this stage as a ground truth or observed source of information. The volunteers are trained to act as sensors to provide real-time information regarding flood depth and location of flood hotspot areas. This geographic information is produced by citizens working as sensors and this type of information is called Volunteered Geographic Information (VGI) (Can et al., 2019; Degrossi et al., 2014; Goodchild, 2007; Goodchild & Li, 2012; Poser et al., 2009; Senaratne et al., 2017).

In order to generate the VGI for the Mumbai floods we have trained local people and the staff of an NGO called Youth for Unity and Voluntary Action (YUVA). By enabling the participation of individuals residing in vulnerable areas, who are often excluded from such studies, this approach not only facilitates the generation of dependable point-based information but also promotes inclusivity. It offers a broader outreach and allows those residing in highly vulnerable regions to contribute to a database that can aid in disaster forecasting and support them in the future. We have trained these volunteers to provide information regarding the location and depth of the flood through WhatsApp and Google Form (Supplementary figure 2). Before every flood event (when there is a heavy rain forecast), the volunteers were informed to stay alert and provide information. For the 2022 monsoon, we gathered the VGI information through WhatsApp for all the extreme rainfall days and used it to validate Twitter data.

## 2.4. Automatic Weather Station data

After the infamous floods of 2005 in Mumbai, the local administration, Municipal Corporation of Greater Mumbai (MCGM), decided to set up a network of Automatic Weather Stations across the city. This provides a dense network of hydro-meteorological information at 15-minute intervals. These stations were installed at the local fire station premises to ensure safety (Ghosh et al., 2021; Singh et al., 2017). Initially, these stations were installed at 26 locations, this number increased to 60 over a period of time, providing a range of meteorological variables, like rainfall, wind speed, wind direction, relative humidity, etc. to MCGM at an interval of 15 minutes.

The rainfall data over these stations were analysed to find trends and patterns in the amount of rainfall and the number of extreme rainfall events. 48 stations were selected which had an uninterrupted data stream of more than 5 years over the period of 2009 - 2022. Rainfall data were first aggregated on a daily scale for the Indian summer monsoon months of June-July-August-September. Wind speeds on the days of such rainfall events were averaged to analyse any significant changes over the years.

## 2.5. Survey

To validate some conclusions drawn from the crowdsourced data, we surveyed a few of the major flood hotspots in Mumbai. These hotspots were chosen after discussing with some local flood experts and officials working on Mumbai floods for the last few years and various news articles. We selected six locations, viz., Dahisar, Borivali, Malad, Khar, Gandhi market, and Hindmata. Due to their lower elevation, Khar, Gandhi market, and Hindmata are the major affected areas among the six chosen locations. To gather more grass root level information, we visited and talked to at least ten people (local shopkeepers or residents) at each of the hotspots with a questionnaire prepared (Supplementary figure 3). We gathered information regarding the severity and reason behind the flood situation and the effectiveness of the mitigation done in the area.

# 3. Results and Discussion

The analysis initially focused on examining the pattern of extreme rainfall events across various regions of Mumbai, using data obtained from AWS. Figure 2(a) presents the spatial variations of annual maxima rainfall averaged over the observation years. We also presented the wind velocity on extreme days (Figure 2(b)) and the elevation map of Mumbai (Figure 2(c)) to understand the reasons behind the extreme rainfall pattern. The spatial patterns of annual maxima in rainfall show higher values in the central regions, with lower elevation, adjacent to the northern high elevation (Figure 2(c)). This is clearly primarily influenced by the orography. The low-lying area receiving high rainfall results in more severe local flooding. The northern regions and the southern tip of the city do not experience such a high rainfall as compared to the central region. The wind plots (Figure 2 (b)) averaged over the extreme days show higher values over the central and in the southern tip of the city. Barring the local influences of orography, by and large; precipitation in central is more than southern tip On July 26, 2005, when Mumbai received very high rainfall, the weather station in central city (Santacruz) received 900 mm of rain in 25 hours, while the southern tip (Colaba) received only 84 mm of rainfall during the same time.

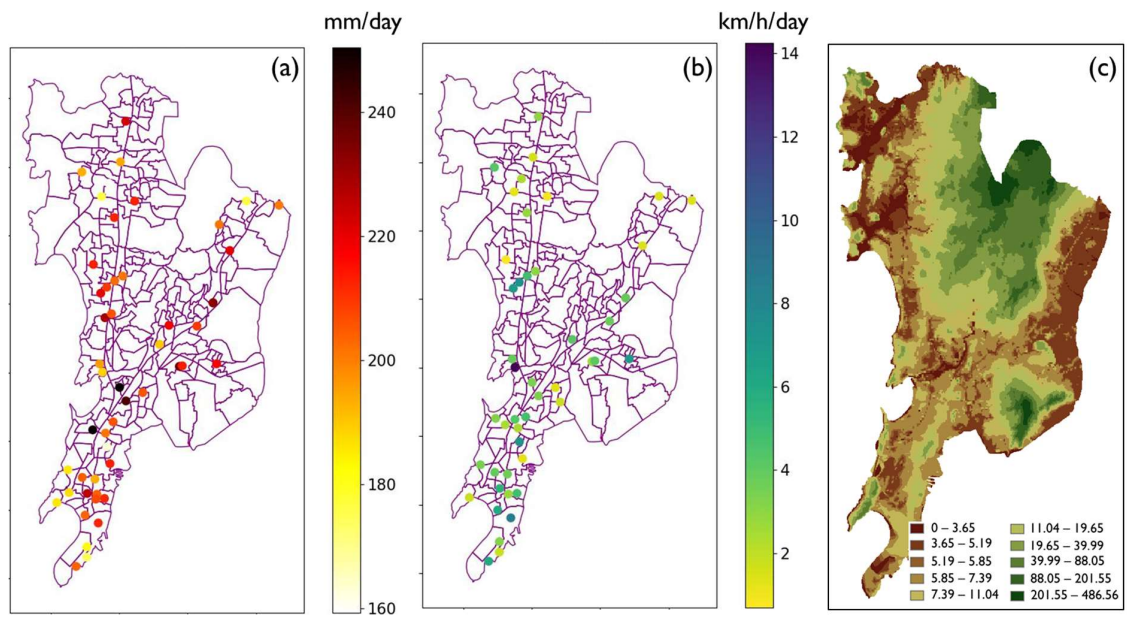

*Figure 2: (a) Average annual maximum rainfall, (b) Average wind on extreme rainfall days;*
*(c) Elevation map for Mumbai*

We first evaluated the usefulness of the crowdsourced flood data for flood data collection and monitoring with a recent extreme event on July 5, 2022. On that day, waterlogging was reported in some parts of the city. The daily accumulated rainfall observed during the day by all the automatic weather stations is shown in Figure 3(a). This rainfall pattern matches with Figure 2 (a), the average extreme day pattern, with the lower elevation central regions receiving higher rainfall and the southern tip receiving comparatively low rainfall. Volunteered geographic information is the most reliable among the other crowdsourced information, but it has spatially and temporally limited availability to provide information for all locations and events. The flood location and depth information was gathered by VGI through WhatsApp from 15 volunteers for the event (Figure 3(b)). Twitter users typically provide more information than VGI; hence, this information can be used to provide flood information at an administrative level to generate flood monitoring maps. We gathered the tweets for the day and filtered them based on location information. We then assigned the useful tweets to the corresponding ward based on location (Figure 3(c)). The entire system is automated. We assumed that the regions with more waterlogging will have more tweets (Bruijn et al., 2019). The tweet map (number of ward-specific tweets) in Figure 3(c) also shows the same pattern as rainfall and VGI, which validates the authenticity of the information. The high inundation regions belong to the low elevation. To scale the elevation for producing the flood potential map, we used the Height Above the Nearest Drainage (HAND) map (Figure 3(d)). We generated the HAND map for Mumbai from a DTM, using the ArcGIS based HAND model (Nobre et al., 2011, 2016; Tripathy et al., 2020)(Supplementary Information). The DTM has a horizontal resolution of 5m and was sourced from the MCGM. The HAND maps cannot be directly converted to or considered as an accurate flood potential map because of the coastal location; however, this map can still serve as a proxy. We found high flooding zones belong to the regions with lower HAND values. This is another proof of the authenticity of the crowdsourced data in Mumbai as a source the flooding information.

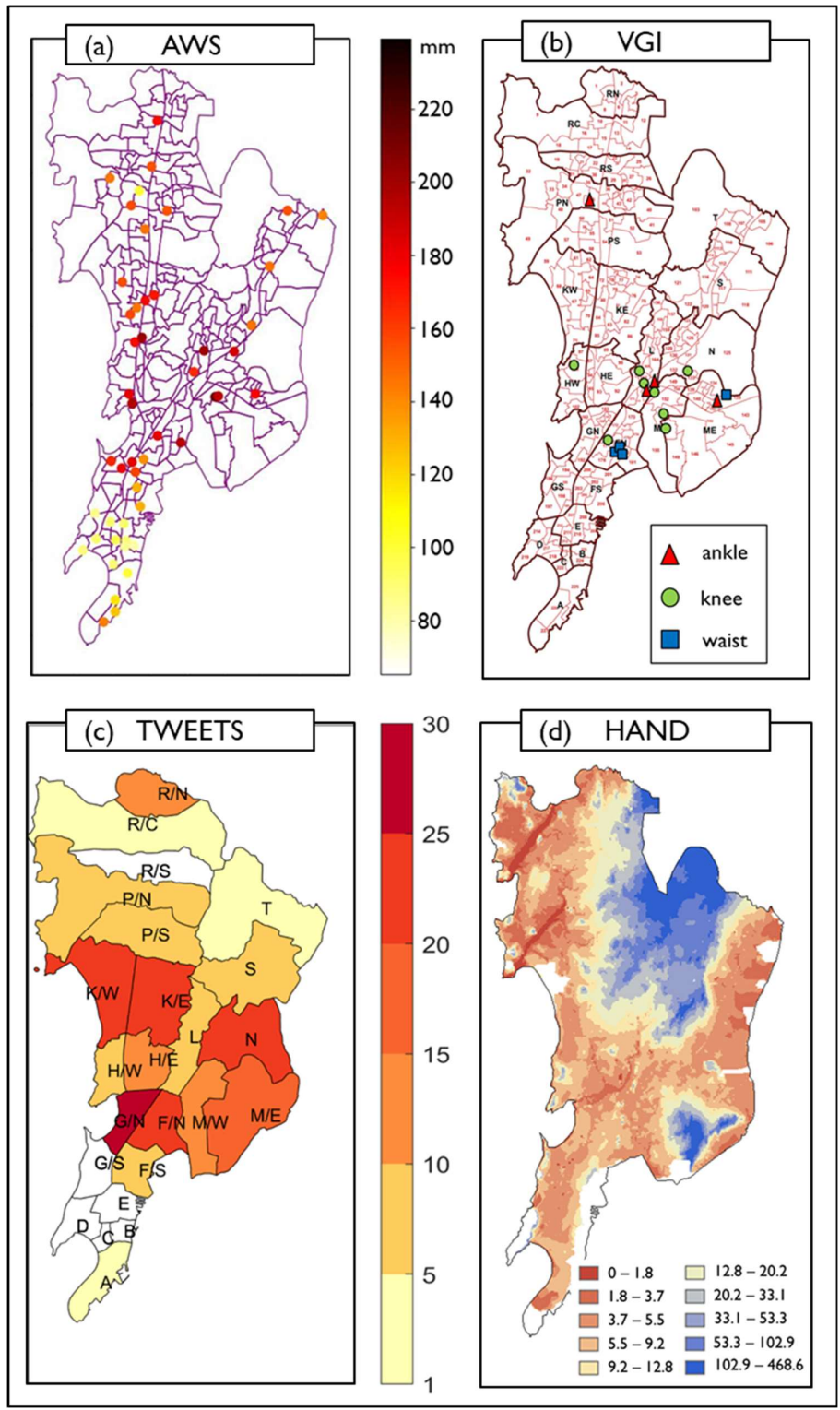

*Figure 3: (a) Observed station based daily accumulated rainfall information, (b) VGI information, and (c) Ward-wise Tweets received, on July 5 2022 (d) HAND value for Mumbai*

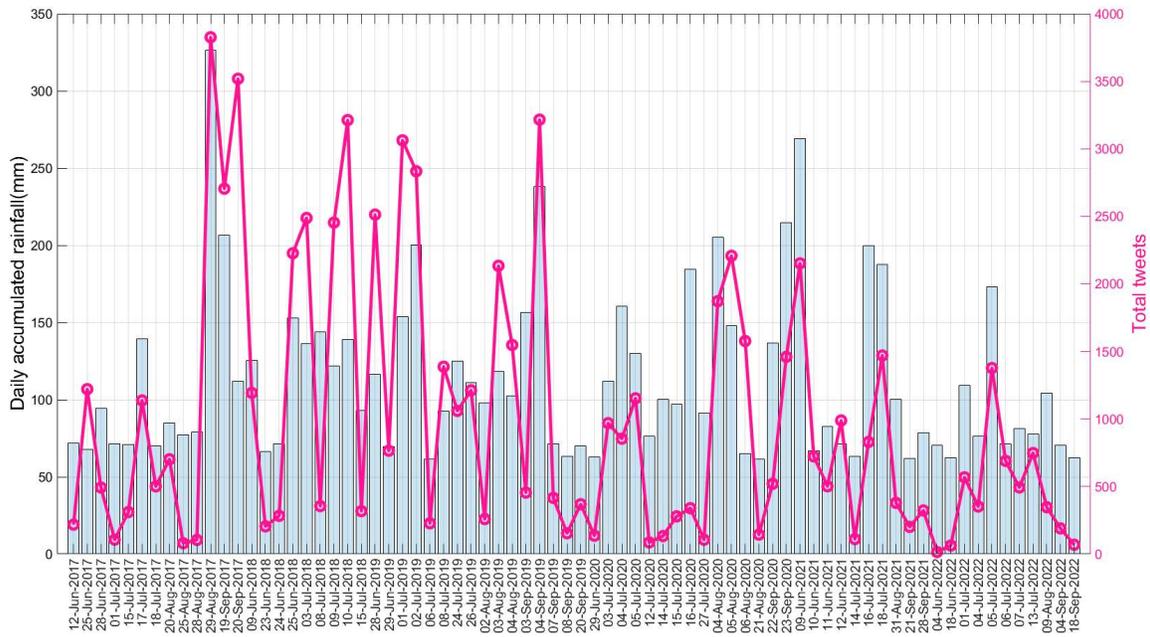

*Figure 4: Daily accumulated rainfall and total number of tweets for past extreme days 2017-2022*

After validating the usefulness of Twitter in correctly identifying an urban flood event in the city, we have further investigated the usefulness of crowdsourced information for past extreme events. Using the observed rainfall information, we have identified the extreme rainfall days during the Indian summer monsoon months of 2017-2022. A peak over the threshold method was employed, setting the threshold of daily accumulated rainfall at 60 mm, to identify these extreme rainfall days. This threshold was chosen based on previous reports of waterlogging in certain areas of the city following rainfall of that magnitude. We identified 75 such extreme days and extracted the tweets for all these days at a time step of 15 minutes (Figure 4). Due to the limitations of the Twitter API, which allows a maximum of 100 tweets to be retrieved per request, we set the time step at 15 minutes to ensure comprehensive retrieval of tweets. The total number of tweets corresponding to the extreme days for 2017-2019 is higher than that of 2020-2022. Contrary to the increase in the total number of Twitter users through the years (*eMarketer, & Insider Intelligence. (December 13, 2022)*) and extreme rainfall events (Mann et al., 2023), the total number of tweets containing flood-related information shows a decreasing trend. This indicates a probable decrease in the water logging situation in the city but does not confirm whether COVID restrictions during 2020 and 2021 played a role. We further investigated the changing patterns of flood scenarios at the flood hotspots through surveys.

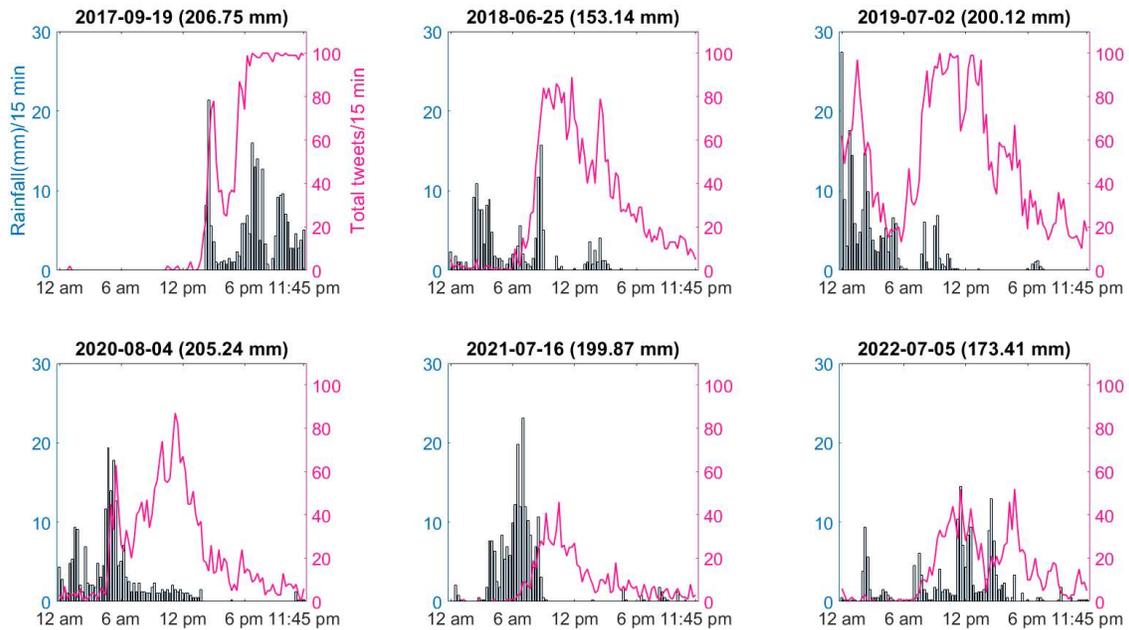

*Figure 5: Extreme events in the years 2017-2021 with rainfall>150mm*

To further analyze the number of tweets for extreme rainfall events across different years we have considered extremely heavy rainfall days (>150 mm/day) from each of the six years 2017-2022. We plotted the observed distribution of rainfall accumulated over 15-minute intervals on each extreme day with the number of tweets received every 15 minutes (Figure 5). For a similar rainfall value, the number of tweets is seen decreasing since 2020. The year 2020 was still affected by the COVID-19 restriction, which may have caused a decrease in the number of tweets. But 2021 and 2022 were post-COVID-19 years, which also show a similar reduction in the number of tweets. A similar trend of decrease in number of tweets can also be seen for extreme rainfall events with a daily accumulated rainfall of around 120-130 mm/day (Supplementary figure 4). This indicates a clear decreasing trend in the reporting of waterlogging in the past years.

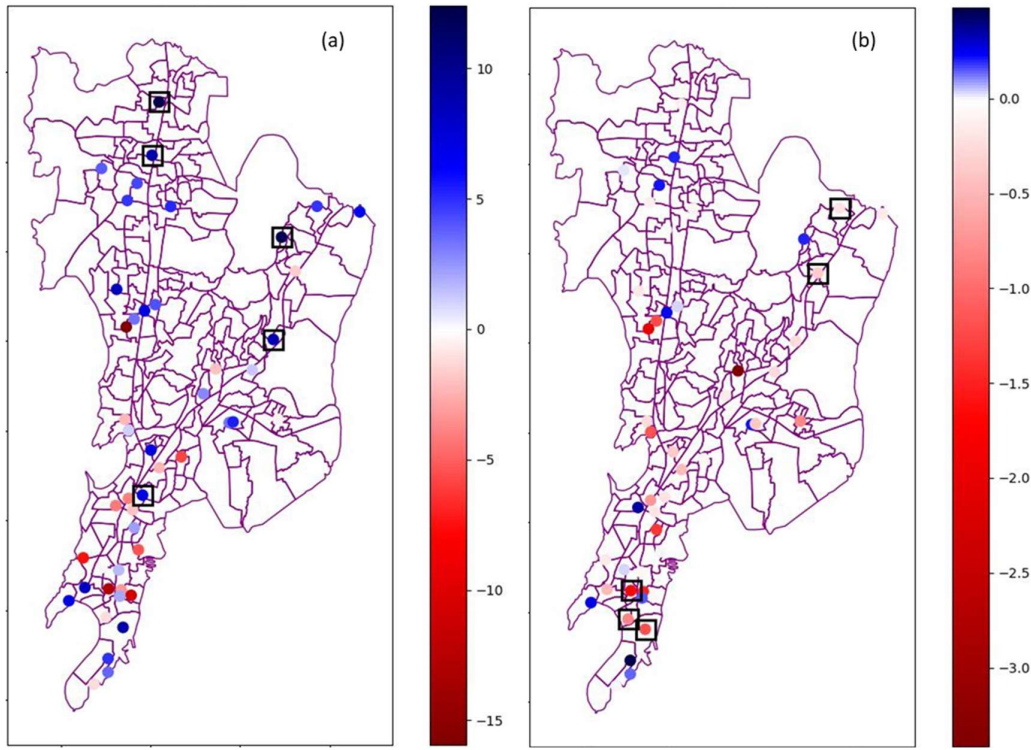

*Figure 6: Trend of (a) Annual maximum rainfall and (b) Average wind speed (stations with statistically significant trends have been marked with a box)*

This decrease in water logging reporting shown in Figures 4 and 5 could be due to a decrease in overall rainfall and extremes in this region or some flood mitigation actions that improved the situation. We first analysed the rainfall trend for Mumbai city to see if any such decreasing trend of extremes exists. The annual maximum rainfall pattern of Mumbai in most of the weather station data shows an increasing trend with a few having statistical significance (Figure 6(a)). We observed increases in the northern and central parts. The average wind speed, on the other hand, did not show any significant change over the years, on the days of annual maximum rainfall. It should be noted that the time period of rainfall trend analysis is longer than the tweet analysis. Hence, it is important to have surveys with the local people to confirm the authenticity of decreasing flooding trends and possible reasons.

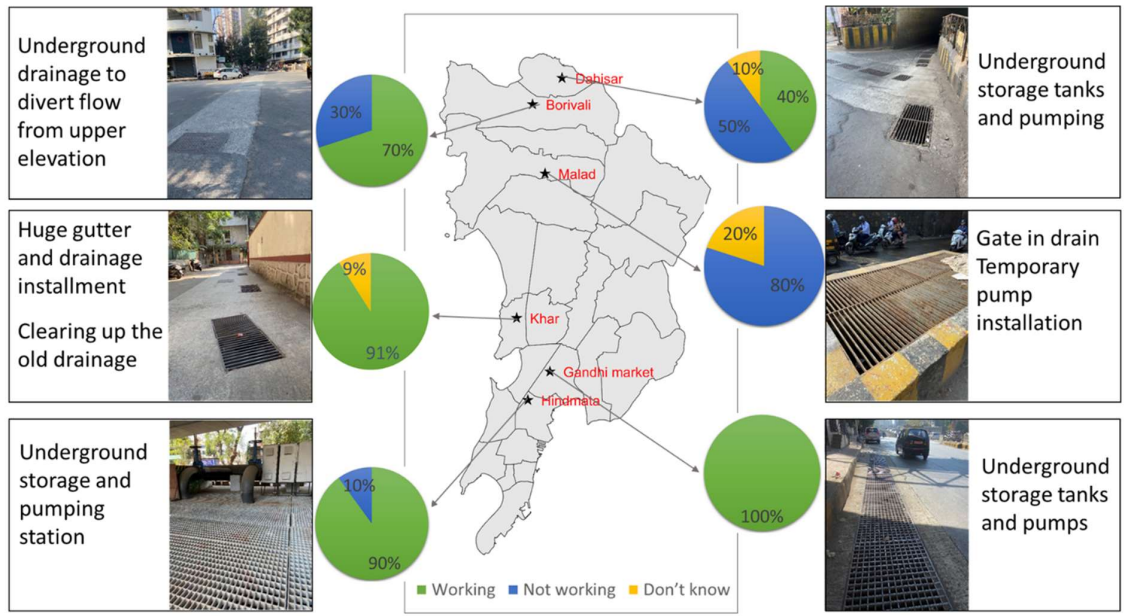

*Figure 7: Survey results from six hotspots in Mumbai*

We conducted a survey of the six flood hotspots in Mumbai, Borivali, Khar, Hindmata, Dahisar, Malad, and Gandhi market. The most common causes of flooding at the hotspots are lower elevation and poor drainage (Supplementary table 2). The BrihanMumbai Municipal Corporation (BMC), the city administration of Mumbai, has taken multiple initiatives in recent years to reduce flooding across the city. Figure 7 presents the mitigation measures taken in the six flood hotspots. We performed the survey to understand if the measures are the reason behind the decrease in flooding. The survey found that most of these mitigations were done in 2020 and 2021. The mitigation measures in the major hotspots like Gandhi market, Hindmata, Khar, and Borivali are working efficiently according to at least 70% of the people surveyed in each location (87% of the total in these 4 locations). The large-scale underground storage tanks and pumping stations are working effectively in reducing the water level significantly with a faster discharge of flood water. However, in the other locations, Dahisar and Malad, we got mixed and negative reviews on the mitigation performed. The reason behind the failure of mitigation in these areas is the location of the nearby drainage at a higher elevation causing the underground tanks to overflow, making them inefficient. This information in the first four locations also agrees with the location-specific decreasing number of tweets (Supplementary figure 4). Such consistency confirms that crowdsourced information can effectively track the flood situation and identify the efficiency of mitigation measures.

# 4. Summary

Existing literature highlights the importance of social media in identifying extremes and generating warnings. Several studies have highlighted the importance of citizens as a source of useful information for analysis and decision-making. Mumbai is a coastal megacity in India that suffers from multiple severe urban flooding events every year during the Indian summer monsoon season. However, it still lacks a reliable flood monitoring and forecasting system that is accessible to the citizens, which could also be beneficial to plan short-term and long-term mitigation strategies. For a densely populated city like Mumbai, the feasibility of using crowdsourced information as a data source for flood forecasting and warning generation is explored here.

In this study, we explored various crowdsourcing datasets to identify those sources which give reliable and authentic information. Twitter is one such crowdsourced data source that is freely available and contains a large database. The default word limit on Twitter makes the tweets shorter in length and loaded with information making it a better choice in comparison to the other social media platforms. The challenge lies in filtering out the incorrect information from the data. Another source is VGI, where trained volunteers work as live sensors to provide real-time information. As a part of this study, we started a campaign earlier in 2022 to train volunteers to work as VGI participants and provide information regarding the flood location and depth through WhatsApp and google forms. The campaign also included open requests to Twitter account holders to become volunteers and provide information by tweeting in a specific format (targeted tweets). Though the data size from VGI is a lot smaller than the Twitter dataset, it is more reliable. The VGI is further used to validate the results obtained through Twitter.

A Python-based automated program was developed to gather information from Twitter. The automated program retrieves tweets for the desired study period. The time interval and keywords are given as inputs and the program saves the Tweets in a tabular (.csv) form. Next, we used filters to identify and gather informative tweets efficiently. Tweets with location information are separated and geotagged (~2-3%). Tweets that contain information on at least one location are termed useful. Locations were searched in a name directory for Mumbai with all possible location names along with the coordinates and the ward that the location belongs to. Each useful tweet is assigned to a ward and a map with ward-wise number of tweets was generated. We found that the map generated with the number of useful tweets is an indicator of the severity, consistent with the VGI and HAND map, a proxy for elevation (Figure 3). The Twitter data can identify lower elevation areas as highly impacted, making it a reliable and granular source of information.

To further validate the usefulness of the crowdsourced data, we analysed all the extreme rainfall days from 2017 to 2022 and found a decrease in the number of reports of waterlogging starting in 2020 (Figure 4 and 5). This was despite the extreme rainfall events not decreasing in most areas. We further validated this by conducting a questionnaire-based survey at some of the major flood hotspots in the city. These surveys inform us about the effectiveness of the mitigation done at these locations by the local city authority (Figure 7).

The present study identifies crowdsourcing as a feasible source of reliable information on floods and waterlogging in Mumbai. The crowdsourcing data is able to identify hotspots and has the potential to generate real-time monitoring. This data can further be used to develop a flood forecasting framework. Crowdsourced information can be very useful in case of the unavailability of fine-resolution observed data sets. These data sets can be used to develop a modern, robust, and complex decision system for more precise and effective decision-making in case of natural disasters. We also propose future works on sentiment analysis and Natural language processing techniques to retrieve more information from the tweets. We expect that citizen participation will only increase if their information is turned into reliable, lifesaving early warning systems.

# References

Arthur, R., Boulton, C. A., Shotton, H., & Williams, H. T. P. (2018). *Social sensing of floods in the UK*. 1–18.

Bielski, C., O'Brien, V., Whitmore, C., Ylinen, K., Juga, I., Nurmi, P., Kilpinen, J., Porras, I., Sole, J. M., Gamez, P., Navarro, M., Alikadic, A., Gobbi, A., Furlanello, C., Zeug, G., Weirathe, M., Martinez, J., Yuste, R., Castro, S., … Rossi, C. (2017). Coupling early warning services, crowdsourcing, and modelling for improved decision support and wildfire emergency management. *2017 IEEE International Conference on Big Data (Big Data)*, 3705–3712. https://doi.org/10.1109/BigData.2017.8258367

Blumberg, A. F., Georgas, N., Yin, L., Herrington, T. O., & Orton, P. M. (2015). Street-Scale Modeling of Storm Surge Inundation along the New Jersey Hudson River Waterfront. *Journal of Atmospheric and Oceanic Technology*, *32*(8), 1486–1497. https://doi.org/10.1175/JTECH-D-14-00213.1

Bondár, I., Steed, R., Roch, J., Bossu, R., Heinloo, A., Saul, J., & Strollo, A. (2020). *Accurate Locations of Felt Earthquakes Using Crowdsource Detections*. *8*(July), 1–9. https://doi.org/10.3389/feart.2020.00272

Bruijn, J. A. De, Moel, H. De, Jongman, B., Ruiter, M. C. De, Wagemaker, J., & Aerts, J. C. J. H. (2019). A global database of historic and real-time flood events based on social media. *Scientific Data*, 1–12. https://doi.org/10.1038/s41597-019-0326-9

Can, R., Kocaman, S., & Gokceoglu, C. (2019). A Convolutional Neural Network Architecture for Auto-Detection of Landslide Photographs to Assess Citizen Science and Volunteered Geographic Information Data Quality. *International Journal of Geo-Information*, *8*(300).

Chen, A. B., Goodall, J. L., Chen, T. D., & Zhang, Z. (2022). Flood resilience through crowdsourced rainfall data collection : Growing engagement faces non-uniform spatial adoption. *Journal of Hydrology*, *609*(November 2021), 127724. https://doi.org/10.1016/j.jhydrol.2022.127724

Çolak, E., & Sunar, F. (2020). The importance of ground-truth and crowdsourcing data for the statistical and spatial analyses of the NASA FIRMS active fires in the Mediterranean Turkish forests. *Remote Sensing Applications: Society and Environment*, *19*, 100327. https://doi.org/https://doi.org/10.1016/j.rsase.2020.100327

de Bruijn, J. A., de Moel, H., Weerts, A. H., de Ruiter, M. C., Basar, E., Eilander, D., & Aerts, J. C. J. H. (2020). Improving the classification of flood tweets with contextual hydrological information in a multimodal neural network. *Computers & Geosciences*, *140*, 104485. https://doi.org/https://doi.org/10.1016/j.cageo.2020.104485

Degrossi, L. C., de Albuquerque, J. P., Fava, M. C., & Mendiondo, E. M. (2014). Flood Citizen Observatory: a crowdsourcing-based approach for flood risk management in Brazil. *SEKE*, 570–575.

Dodman, D., B. Hayward, M. Pelling, V. Castan Broto, W. Chow, E. Chu, R. Dawson, L. Khirfan, T. McPhearson, A. Prakash, Y. Zheng, and G. Ziervogel, 2022: Cities, Settlements, and Key Infrastructure. In: Climate Change 2022: Impacts, Adaptation, and Vulnerability. Contribution of Working Group II to the Sixth Assessment Report of the Intergovernmental Panel on Climate Change [H.-O. Pörtner, D.C. Roberts, M. Tignor,

E.S. Poloczanska, K. Mintenbeck, A. Alegría, M. Craig, S. Langsdorf, S. Löschke, V. Möller, A. Okem, B. Rama (eds.)]. Cambridge University Press, Cambridge, UK and New York, NY, USA, pp. 907–1040, doi:10.1017/9781009325844.008.

eMarketer, & Insider Intelligence. (December 13, 2022). *Number of Twitter users worldwide from 2019 to 2024 (in millions) [Graph]. In Statista. Retrieved January 30, 2023, from https://www.statista.com/statistics/303681/twitter-users-worldwide*. (n.d.).

Flanagin, A. J., & Metzger, Æ. M. J. (2008). *The credibility of volunteered geographic information*. *July*, 137–148. https://doi.org/10.1007/s10708-008-9188-y

Gebremedhin, E. T., Basco-Carrera, L., Jonoski, A., Iliffe, M., & Winsemius, H. (2020). Crowdsourcing and interactive modelling for urban flood management. *Journal of Flood Risk Management*, *13*(2), 1–15. https://doi.org/10.1111/jfr3.12602

Ghosh, M., Singh, J., Sekharan, S., Ghosh, S., Zope, P. E., & Karmakar, S. (2021). Rationalization of automatic weather stations network over a coastal urban catchment: A multivariate approach. *Atmospheric Research*, *v. 254*(2021), 105511--2021 v.254. https://doi.org/10.1016/j.atmosres.2021.105511

Goodchild, M. F. (2007). *Citizens as sensors : the world of volunteered geography*. *November*, 211–221. https://doi.org/10.1007/s10708-007-9111-y

Goodchild, M. F., & Li, L. (2012). Assuring the quality of volunteered geographic information. *Spatial Statistics*, *1*, 110–120. https://doi.org/https://doi.org/10.1016/j.spasta.2012.03.002

Joshua Roesslein. (2009). *Tweepy documentation*. http://tweepy. readthedocs. io/en/v3, 5.

Hallegatte, S., Ranger, N., Bhattacharya, S., Bachu, M., Priya, S., Dhore, K., Rafique, F., Mathur, P., Naville, N., Henriet, F., Patwardhan, A., Narayanan, K., Ghosh, S., Karmakar, S., Patnaik, U., Abhayankar, A., Pohit, S., Corfee-Morlot, J., & Herweijer, C. (2010). *Flood Risks, Climate Change Impacts and Adaptation Benefits in Mumbai*. *27*. https://doi.org/https://doi.org/10.1787/5km4hv6wb434-en

Havas, C., Resch, B., Francalanci, C., Id, B. P., Scalia, G., Fernandez-marquez, J. L., Achte, T. Van, Zeug, G., Rosa, M., Mondardini, R., Grandoni, D., Kirsch, B., Kalas, M., Lorini, V., & Rüping, S. (2017). *E2mC : Improving Emergency Management Service Practice through Social Media and Crowdsourcing Analysis in Near Real Time*. https://doi.org/10.3390/s17122766

Helmrich, A. M., Ruddell, B. L., Bessem, K., Chester, M. V., Chohan, N., Doerry, E., Eppinger, J., Garcia, M., Goodall, J. L., Lowry, C., & Zahura, F. T. (2021). Opportunities for crowdsourcing in urban flood monitoring. *Environmental Modelling and Software*, *143*, 105124. https://doi.org/10.1016/j.envsoft.2021.105124

Hultquist, C., & Cervone, G. (2020). Integration of Crowdsourced Images, USGS Networks, Remote Sensing, and a Model to Assess Flood Depth during Hurricane Florence. *Remote Sensing*, *12*(5). https://doi.org/10.3390/rs12050834

Kim, Y., Eisenberg, D. A., Bondank, E. N., Chester, M. V, Mascaro, G., & Underwood, B. S. (2017). Fail-safe and safe-to-fail adaptation: decision-making for urban flooding under climate change. *Climatic Change*, *145*(3), 397–412. https://doi.org/10.1007/s10584-017-2090-1

Kosmala, M., Wiggins, A., Swanson, A., & Simmons, B. (2016). Assessing data quality in citizen science. *Frontiers in Ecology and the Environment*, *14*(10), 551–560. https://doi.org/https://doi.org/10.1002/fee.1436

Minson, S. E., Brooks, B. A., Glennie, C. L., Murray, J. R., Langbein, J. O., Owen, S. E., Heaton, T. H., Iannucci, R. A., & Hauser, D. L. (2015). *Crowdsourced earthquake early warning*. April, 1–8.

Nied, M., Pardowitz, T., Nissen, K., Ulbrich, U., Hundecha, Y., & Merz, B. (2014). On the relationship between hydro-meteorological patterns and flood types. *Journal of Hydrology*, *519*, 3249–3262. https://doi.org/https://doi.org/10.1016/j.jhydrol.2014.09.089

Nobre, A D, Cuartas, L. A., Hodnett, M., Rennó, C. D., Rodrigues, G., Silveira, A., & Waterloo, M. (2011). Height Above the Nearest Drainage – a hydrologically relevant new terrain model. *Journal of Hydrology*, *404*(1–2), 13–29. https://doi.org/10.1016/j.jhydrol.2011.03.051

Nobre, Antonio Donato, Cuartas, L. A., Momo, M. R., Severo, D. L., Pinheiro, A., & Nobre, C. A. (2016). HAND contour: A new proxy predictor of inundation extent. *Hydrological Processes*, *30*(2), 320–333. https://doi.org/10.1002/hyp.10581

Oliveira, A. C. M., Botega, L. C., Saran, J. F., Silva, J. N., Melo, J. O. S. F., Tavares, M. F. D., & Neris, V. P. A. (2019). Crowdsourcing, data and information fusion and situation awareness for emergency Management of forest fires: The project DF100Fogo (FDWithoutFire). *Computers, Environment and Urban Systems*, *77*, 101172. https://doi.org/https://doi.org/10.1016/j.compenvurbsys.2017.08.006

Paul, J. D., Buytaert, W., Allen, S., Ballesteros-Cánovas, J. A., Bhusal, J., Cieslik, K., Clark, J., Dugar, S., Hannah, D. M., Stoffel, M., Dewulf, A., Dhital, M. R., Liu, W., Nayaval, J. L., Neupane, B., Schiller, A., Smith, P. J., & Supper, R. (2018). Citizen science for hydrological risk reduction and resilience building. *WIREs Water*, *5*(1), e1262. https://doi.org/https://doi.org/10.1002/wat2.1262

Poser, K., Kreibich, H., & Dransch, D. (2009). Assessing Volunteered Geographic Information for Rapid Flood Damage Estimation. *12th AGILE International Conference on Geographic Information Science 2009 Leibniz Universität Hannover, Germany*, 1–9.

Roberts, A. D. (2022). Comparison Of Satellite-Correlated Crowdsourced And Geostatistical Temperature Projections For An Atlanta Heatwave Daily Snapshot : How Differences In Spatially-Variable Predictions May Be Correlated To Economic Status And Land Cover / Land Use. *Sustainable Futures*, *4*(December 2021), 100078. https://doi.org/10.1016/j.sftr.2022.100078

Rosenzweig, B. R., Herreros Cantis, P., Kim, Y., Cohn, A., Grove, K., Brock, J., Yesuf, J., Mistry, P., Welty, C., McPhearson, T., Sauer, J., & Chang, H. (2021). The Value of Urban Flood Modeling. *Earth's Future*, *9*(1). https://doi.org/10.1029/2020EF001739

See, L. (2019). A review of citizen science and crowdsourcing in applications of pluvial flooding. *Frontiers in Earth Science*, *7*(March), 1–7. https://doi.org/10.3389/feart.2019.00044

Senaratne, H., Mobasheri, A., Ali, A. L., Capineri, C., & Haklay, M. M. (2017). A review of volunteered geographic information quality assessment methods. *International Journal of Geographical Information Science*, *31*(1), 139–167. https://doi.org/10.1080/13658816.2016.1189556

Joshua Roesslein. (2009). *Tweepy documentation*.

Mann, R., Gupta, A., Dhorde, A., & Sharma, S. (2023). Observed trends and coherent changes in daily rainfall extremes over Greater Mumbai, 1985–2020. *Theoretical and Applied Climatology*, *151*(3), 1889–1910. https://doi.org/10.1007/s00704-022-04354-4

Shastri, H., Barik, B., Ghosh, S., Venkataraman, C., & Sadavarte, P. (2017). Flip flop of Day-night and Summer-Winter Surface Urban Heat Island Intensity in India. *Scientific Reports*, *7*(1), 40178. https://doi.org/10.1038/srep40178

Singh, J., Sekharan, S., Karmakar, S., Ghosh, S., Zope, P. E., & Eldho, T. I. (2017). Spatio-temporal analysis of sub-hourly rainfall over Mumbai , India : Is statistical forecasting futile ? *Journal of Earth System Science*, *126*(38), 1–15. https://doi.org/10.1007/s12040-017-0817-z

Smith, L., Liang, Q., James, P., & Lin, W. (2017). Assessing the utility of social media as a data source for flood risk management using a real-time modelling framework. *Journal of Flood Risk Management*, *10*, 370–380. https://doi.org/10.1111/jfr3.12154

Songchon, C., Wright, G., & Beevers, L. (2021). Computers , Environment and Urban Systems Quality assessment of crowdsourced social media data for urban flood management. *Computers, Environment and Urban Systems*, *90*(August), 101690. https://doi.org/10.1016/j.compenvurbsys.2021.101690

Sy, B., Frischknecht, C., Dao, H., Consuegra, D., & Giuliani, G. (2019). Flood hazard assessment and the role of citizen science. *Journal of Flood Risk Management*, *12*(S2), e12519. https://doi.org/https://doi.org/10.1111/jfr3.12519

Tavra, M., Racetin, I., & Peroš, J. (2021). The role of crowdsourcing and social media in crisis mapping: a case study of a wildfire reaching Croatian City of Split. *Geoenvironmental Disasters*, *8*(1). https://doi.org/10.1186/s40677-021-00181-3

Tripathy, S. S., Vittal, H., Karmakar, S., & Ghosh, S. (2020). Flood risk forecasting at weather to medium range incorporating weather model , topography , socio-economic information and land use exposure. *Advances in Water Resources*, *146*(October), 103785. https://doi.org/10.1016/j.advwatres.2020.103785

Venter, Z. S., Chakraborty, T., & Lee, X. (2021). *Crowdsourced air temperatures contrast satellite measures of the urban heat island and its mechanisms*. 1–10.

Wang, R., Mao, H., Wang, Y., Rae, C., & Shaw, W. (2018). Computers and Geosciences Hyper-resolution monitoring of urban fl ooding with social media and crowdsourcing data. *Computers and Geosciences*, *111*(September 2017), 139–147. https://doi.org/10.1016/j.cageo.2017.11.008

Wang, Y., Chen, A. S., Fu, G., Djordjević, S., Zhang, C., & Savić, D. A. (2018). An integrated framework for high-resolution urban flood modelling considering multiple information sources and urban features. *Environmental Modelling & Software*, *107*, 85–95. https://doi.org/https://doi.org/10.1016/j.envsoft.2018.06.010

Winsemius, H. C., Aerts, J. C. J. H., Beek, L. P. H. Van, Bierkens, M. F. P., Bouwman, A., Jongman, B., Kwadijk, J. C. J., Ligtvoet, W., Lucas, P. L., Vuuren, D. P. Van, & Ward, P. J. (2016). Global drivers of future river flood risk. *Nature Climate Change*, *6*(April), 381–385. https://doi.org/10.1038/NCLIMATE2893

Witherow, M. A., Elbakary, M. I., Iftekharuddin, K. M., & Cetin, M. (2018). Analysis of Crowdsourced Images for Flooding Detection. In J. M. R. S. Tavares & R. M. Natal Jorge (Eds.), *VipIMAGE 2017* (pp. 140–149). Springer International Publishing.

Supplementary Information

# Analysis of Mumbai Floods in recent Years with Crowdsourced Data

Shrabani Sailaja Tripathy[1], Sautrik Chaudhuri[2], Raghu Murtugudde[1], Vedant Mharte[3], Dulari Parmar[4], Manasi Pinto[4], P.E. Zope[2], Vishal Dixit[1], Subimal Ghosh[1,2,*]

[1]Interdisciplinary Program in Climate Studies, Indian Institute of Technology Bombay, Powai, Mumbai-400076, India.
[2]Department of Civil Engineering, Indian Institute of Technology Bombay, Powai, Mumbai-400076, India.
[3]Electronics and Telecommunication Engineering, Vidyalankar Institute of Technology, Mumbai 400 037, India
[4]Youth for Unity and Voluntary Action (YUVA), Kharghar, Navi Mumbai, 410201, India

*Corresponding author, subimal@iitb.ac.in

**Supplementary table 1:** Example of tweets extracted

| | |
|---|---|
| Constituency of Ex-Mayor of **Vasai-Virar** already under water! <br> #VasaiVirar <br> #MumbaiRains https://t.co/2g1OQpPAKu | 2022-07-01 <br> 12:53:25+00:00 |
| Heavy waterlogging in **Ghatkopar East, Tilak Nagar and Kurla West**. <br> To track #MumbaiRains and waterlogging, download #MumbaiRainApp now: https://t.co/Edh1eol7U5 <br> https://t.co/R1fKb7YGLE | 2022-07-01 <br> 13:22:18+00:00 |
| Andheri subway has been shut at 640 due to water clogging <br> Traffic diverted to **Gokhale bridge** for east west connectivity <br> DN Nagar traffic division managing the traffic flow <br> @mumbaitraffic #mumbairains | 2022-07-01 <br> 13:16:35+00:00 |
| **Andheri Subway** water logged, please avoid that route. <br> They closed it just now | 2022-07-01 <br> 13:09:41+00:00 |
| @MumbaiPolice #MumbaiRains #mumbaiupdate @IndiaWeatherMan #avoidandherisubway | |
| Knee level flooding in half hour. Drains not working? #mumbairains am I not supposed to ever get in and out of or lane? | 2022-07-01 12:01:58+00:00 |
| Water logging is clear at **Hindamata Junction**, traffic is moving. #MTPTrafficUpdate #MumbaiRains @MumbaiPolice | 2022-07-12 05:35:56+00:00 |

**Supplementary table 2:** Major hotspots surveyed and the reason for flooding in these areas

| Places visited | Reason for flooding |
|---|---|
| D. N.Mhatre road, Borivali | 1. Low elevation of the road causes flow from the high elevation areas |
| Dahisar Subway | 1. Low elevation<br>2. Nearby open gutter/drain |
| Malad Subway | 1. Low elevation<br>2. Nearby big drain |
| Gandhi market | 1. Low elevation |
| Hindmata | 1. Low elevation |
| Jaybharat Society area in Khar | 1. Low elevation<br>2. Poor drainage<br>3. Poorly planned construction |

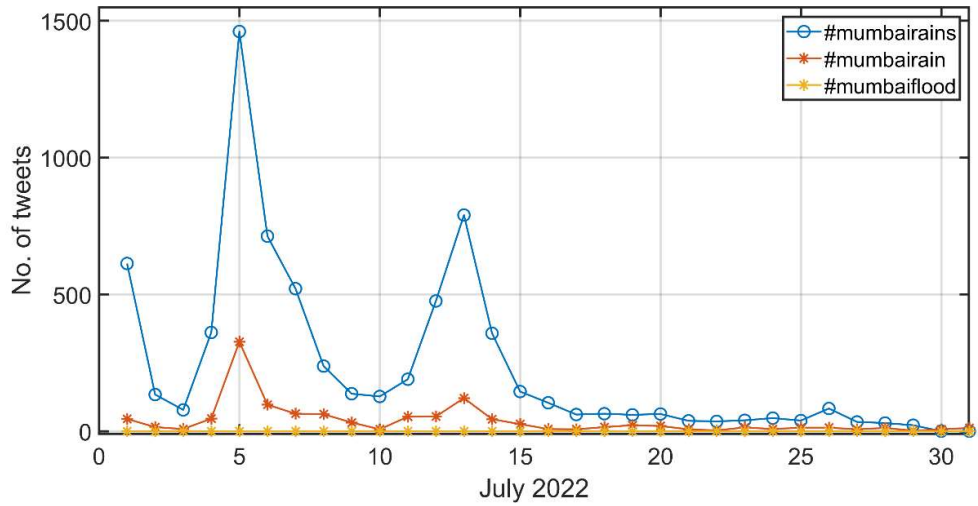

**Supplementary figure 1**: The number of tweets recorded daily for three hashtags

(#mumbairains, #mumbairain, and #mumbaiflood) for July 2022

**Supplementary figure 2:** Training volunteers to gather information for VGI

## Questionnaire

Area: _______________________________________________

Name: _______________________________________________

How long have you been staying/working here? _______________________________

How severe is the flood situation in this location?

   a. Very
   b. Moderate
   c. Low

What is the reason behind the flood in this area? _______________________________

Was the flood situation better this year or last year as compared to the previous years?

   a. Yes
   b. No
   c. Don't know

Did BMC do any mitigation in this area?

   a. Yes
   b. No

If yes, when?

   a. 2020
   b. 2021
   c. 2022

What mitigation practice was done here? _______________________________

Do you think this mitigation working?

   a. Yes
   b. No
   c. Don't know

**Supplementary figure 3:** Questionnaire used to perform surveys regarding the flood situation at the hotspot locations

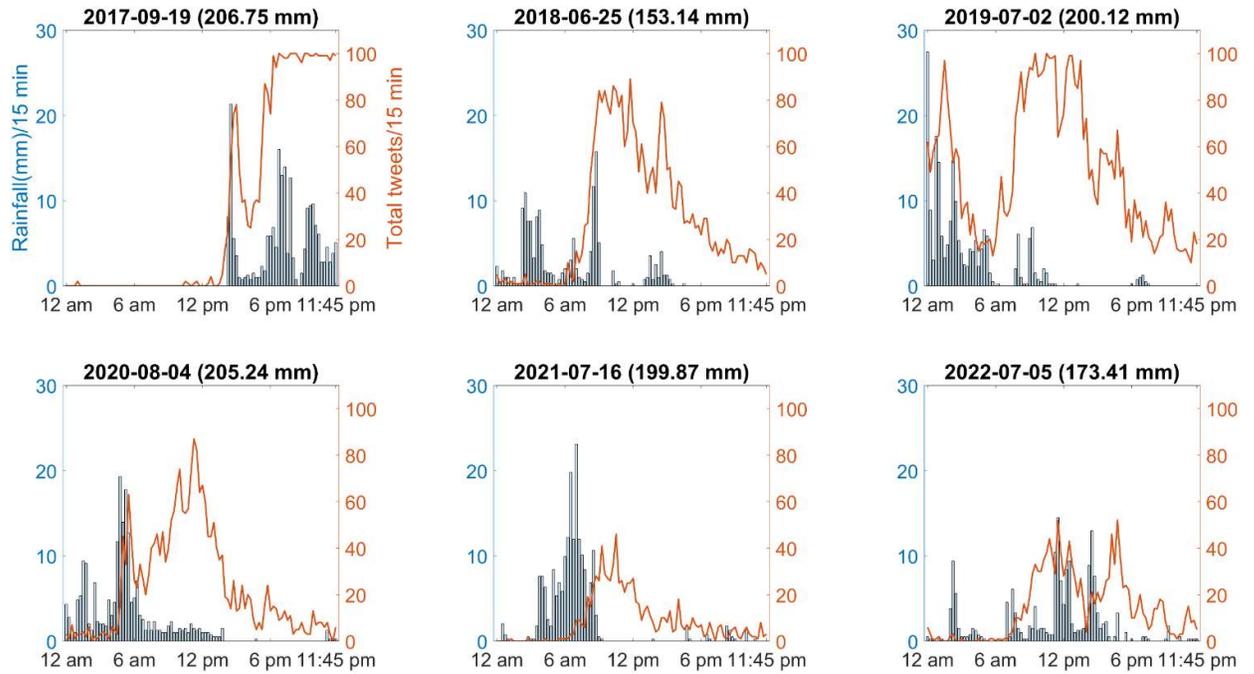

**Supplementary figure 4:** Extreme events in the years 2017-2022 with rainfall 120-130 mm. The blue histogram shows the accumulated rainfall (mm/15 min) and the orange line shows the total number of tweets received every 15 minutes.

**Supplementary figure 5:** Location-wise relative number of tweets for all the extreme events

$$\left( \frac{\text{Total useful tweet in area 'i' in year 'n'}}{\text{Total useful tweet in year 'n'}} \right)$$

# Height Above the nearest drainage

Height Above the Nearest Drainage (HAND) is commonly used as a proxy for elevation (Nobre et al., 2011; Nobre et al., 2016; Rahmati et al., 2018; Rennó et al., 2008; Tripathy et al., 2020), especially in the context of hydrological analysis. The HAND model enables better hydrological analysis by incorporating the influence of drainage patterns. It helps identify areas of higher elevation that contribute to the flow of water, providing insights into the drainage network and potential flood-prone areas (Garousi-Nejad et al., 2019; Johnson et al., 2019; Li & Demir, 2023). This approach involves normalizing Digital Terrain Model (DTM) values by converting the elevation relative to sea level into elevation relative to the nearest drainage. For the generation of the comprehensive HAND map of Mumbai, we used an ArcGIS-based program developed by Nobre et al. (2011).

First, a DTM with sinks is used as input in the HAND model (Supplementary figure 6(a)). These sinks, which are depressions or areas with lower elevation than their surroundings, are then connected using a depression breaching method (Martz & Garbrecht, 1999) to create a hydrologically coherent DTM (Supplementary figure 6(c)). Next, flow directions are determined using the hydrologically coherent DTM using the D8 method (Tarboron, 1997). This method assigns the flow from each grid to its steepest downslope neighbour (Supplementary figure 6(d)). This information is used to compute flow accumulation, which represents the number of grids draining into each grid.

To generate a drainage network map, a threshold for flow accumulation is applied (Supplementary figure 6(e)). Grids with accumulated flow above the threshold (set as 10 in the example) are considered to be a part of the drainage network (Supplementary figure 6(f)). The threshold can be implemented either manually or automatically by utilizing a precisely mapped stream network (MSN) as input, which provides an accurate drainage network map. By analysing the drainage network and flow directions, the nearest drainage grid for each grid is identified.

In the illustration given in Supplementary figure 6(g), each drainage grid is assigned a specific colour, and the corresponding non-drainage grids are depicted with a lighter shade of the same colour. The HAND value for each grid (Supplementary figure 6(h)) is then calculated as the difference between its elevation and the elevation of the nearest drainage grid. This provides a measure of the height above the nearest drainage for each location.

The HAND method allows for a more nuanced assessment of elevation, as it considers both topography and proximity to drainage. It helps identify areas that are more susceptible to flooding and provides valuable information for flood risk management and mitigation strategies.

Overall, the HAND method is a valuable tool to represent the local terrain and has been widely used in flood-related studies. It enhances our understanding of the relationship between elevation and drainage, providing effective visualization and interpretation, valuable insights for flood forecasting, hazard assessment, and land use planning. The HAND model can be applied to various geographical regions and scales, making it a versatile tool for studying and analyzing different landscapes. It can be used in both urban and rural areas to assess flood risk, inform land management decisions, and support disaster preparedness and response efforts.

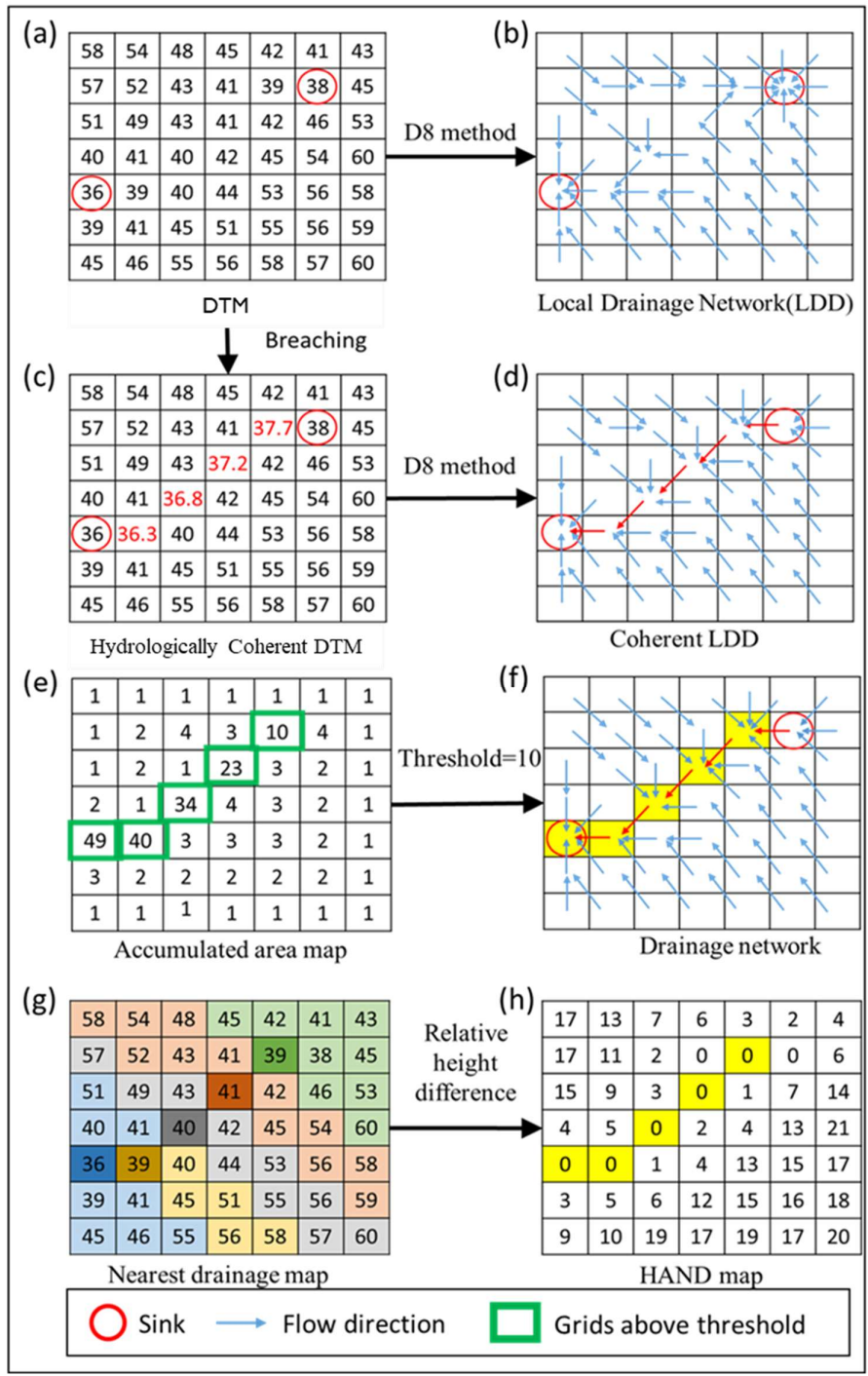

**Supplementary figure 6:** Demonstration of HAND method: (a) Example DTM, (b) Original LDD, (c) Hydrologically coherent DTM, (d) coherent LDD, (e) Accumulated area map, (f) Drainage network, (g) Nearest drainage map, and (h) HAND map.

# References

Garousi-Nejad, I., Tarboton, D. G., Aboutalebi, M., & Torres-Rua, A. F. (2019). Terrain Analysis Enhancements to the Height Above Nearest Drainage Flood Inundation Mapping Method. *Water Resources Research*, *55*(10), 7983–8009. https://doi.org/https://doi.org/10.1029/2019WR024837

Johnson, J. M., Munasinghe, D., Eyelade, D., & Cohen, S. (2019). An integrated evaluation of the National Water Model (NWM)--Height Above Nearest Drainage (HAND) flood mapping methodology. *Natural Hazards and Earth System Sciences*, *19*(11), 2405–2420. https://doi.org/10.5194/nhess-19-2405-2019

Li, Z., & Demir, I. (2023). U-net-based semantic classification for flood extent extraction using SAR imagery and GEE platform: A case study for 2019 central US flooding. *Science of The Total Environment*, *869*, 161757. https://doi.org/https://doi.org/10.1016/j.scitotenv.2023.161757

Martz, L. W., & Garbrecht, J. (1999). An outlet breaching algorithm for the treatment of closed depressions in a raster DEM. *Computers and Geosciences*, *25*(7), 835–844. https://doi.org/10.1016/S0098-3004(99)00018-7

Nobre, A. D., Cuartas, L. A., Hodnett, M., Rennó, C. D., Rodrigues, G., Silveira, A., Waterloo, M., & Saleska, S. (2011). Height Above the Nearest Drainage - a hydrologically relevant new terrain model. *Journal of Hydrology*, *404*(1–2), 13–29. https://doi.org/10.1016/j.jhydrol.2011.03.051

Nobre, Antonio Donato, Cuartas, L. A., Momo, M. R., Severo, D. L., Pinheiro, A., & Nobre, C. A. (2016). HAND contour : a new proxy predictor of inundation extent. *Hydrological Processes*, *333*(August 2015), 320–333. https://doi.org/10.1002/hyp.10581

Rahmati, O., Kornejady, A., Samadi, M., Nobre, A. D., & Melesse, A. M. (2018). Development of an automated GIS tool for reproducing the HAND terrain model. *Environmental Modelling and Software*, *102*, 1–12. https://doi.org/10.1016/j.envsoft.2018.01.004

Rennó, C. D., Nobre, A. D., Cuartas, L. A., Soares, J. V., Hodnett, M. G., Tomasella, J., & Waterloo, M. J. (2008). HAND, a new terrain descriptor using SRTM-DEM: Mapping terra-firme rainforest environments in Amazonia. *Remote Sensing of Environment*, *112*(9), 3469–3481. https://doi.org/10.1016/j.rse.2008.03.018

Tarboron, G. (1997). A new method for the determination of flow directions and upslope areas in grid digital elevation models. *Water Resources Research*, *33*(2), 309–319.

Tripathy, S. S., Vittal, H., Karmakar, S., & Ghosh, S. (2020). Flood risk forecasting at weather to medium range incorporating weather model , topography , socio-economic information and land use exposure. *Advances in Water Resources*, *146*(October), 103785. https://doi.org/10.1016/j.advwatres.2020.103785